\documentclass[notitlepage,superscriptaddress,twocolumn]{revtex4-2}

\usepackage{amsfonts,amsmath,amssymb,stmaryrd}

\usepackage{graphicx}

\usepackage{siunitx}
\usepackage[hidelinks]{hyperref}
\usepackage[nameinlink]{cleveref}
\Crefname{equation}{Eq.}{Eqs.}
\Crefname{figure}{Fig.}{Figs.}
\Crefname{tabular}{Tab.}{Tabs.}
\Crefname{table}{Tab.}{Tabs.}

\usepackage{nicefrac}
\usepackage[autostyle=true]{csquotes}
\usepackage{tikz}
\usepackage{tabularx}
\usepackage{todonotes}
\usepackage{bm}

\newcommand{\boltzmann}{k_\mathrm{B}}
\newcommand{\bv}{\mathbf}
\newcommand{\iu}{{i\mkern1mu}}
\newcommand{\avg}[1]{\langle #1 \rangle}

\newcommand{\tbreakdown}{t^\mathrm{h}_\mathrm{on}}
\newcommand{\treoccurrence}{t^\mathrm{h}_\mathrm{off}}
\newcommand{\trise}{t^\mathrm{d}_\mathrm{on}}
\newcommand{\tdecay}{t^\mathrm{d}_\mathrm{off}}

\newcommand{\tc}{T_\mathrm{c}}

\DeclareSIUnit\gauss{G}
\DeclareSIUnit\bohrradii{a_0}

\begin{document}
	\title{Observing the loss and revival of long-range phase coherence through disorder quenches}
	
	\author{Benjamin Nagler}
	\affiliation{Department of Physics and Research Center OPTIMAS, Technische Universit\"at Kaiserslautern, Germany}
	
	\author{Sian Barbosa}
	\affiliation{Department of Physics and Research Center OPTIMAS, Technische Universit\"at Kaiserslautern, 67663 Kaiserslautern, Germany}
	
	\author{Jennifer Koch}
	\affiliation{Department of Physics and Research Center OPTIMAS, Technische Universit\"at Kaiserslautern, 67663 Kaiserslautern, Germany}
	
	\author{Giuliano Orso}
	\email{email: giuliano.orso@u-paris.fr}
	\affiliation{Université de Paris, Laboratoire Matériaux et Phénomènes Quantiques, CNRS, F-75013 Paris, France}
	
	\author{Artur Widera}
	\email{email: widera@physik.uni-kl.de}
	\affiliation{Department of Physics and Research Center OPTIMAS, Technische Universit\"at Kaiserslautern, 67663 Kaiserslautern, Germany}
	
	\date{\today}
	
	\begin{abstract}
		Relaxation of quantum systems is a central problem in nonequilibrium physics. In contrast to classical systems, the underlying quantum dynamics results not only from atomic interactions but also from the long-range coherence of the many-body wave function.
		Experimentally, nonequilibrium states of quantum fluids are usually created using moving objects or laser potentials, directly perturbing and detecting the system's density. 
		However the fate of long-range phase coherence for hydrodynamic motion of disordered quantum systems is less explored, especially in three dimension.
		Here, we unravel how the density and phase coherence of a Bose-Einstein condensate of $^6$Li$_2$ molecules respond upon quenching on or off an optical speckle potential. 
		We find that, as the disorder is switched on, long-range phase coherence breaks down one order of magnitude faster than the density of the quantum gas responds. After removing it, the system needs two orders of magnitude longer times to reestablish quantum coherence, compared to the density response. 
		We compare our results with numerical simulations of the Gross-Pitaevskii equation on large three-dimensional grids, finding an overall good agreement.
		Our results shed light on the importance of long-range coherence and possibly long-lived phase excitations for the relaxation of nonequilibrium quantum many-body systems. 
	\end{abstract}
	
	\maketitle

	Macroscopic quantum phenomena such as superconductivity and superfluidity are central to our understanding of many-body quantum systems and play an important role in emerging quantum technologies~\cite{Degen2017}. Their fascinating properties are tightly linked to the existence of a global wave function 
	\begin{align}	
		\psi = \sqrt{n} e^{\iu \phi}, 
	\end{align}
	with $n$ being the density and $\phi$ the quantum phase.  Long-range phase coherence, i.e., a fixed phase relation between far distant locations in the quantum system, is crucial for establishing superfluid properties in interacting systems~\cite{Devreese2012}. Microscopically, a large number of particles occupy the same quantum state phase-coherently, as first recognized by Fritz London providing a description of the properties of superfluid $^4$He~\cite{Gavroglu1988}, which has been successfully applied to the theoretical understanding and experimental control of Bose-Einstein condensates (BEC) in dilute atomic gases~\cite{Ketterle2008}. 
	The macroscopic quantum phase $\phi$ has been revealed in numerous interference experiments on BECs, including measurements of the first-order correlation function~\cite{Andrews1997, Hagley1999, Bloch2000, Chin06}, its dynamics~\cite{Hadzibabic2006, Ritter2007, Hofferberth07} in low-dimensional gases, or its statistics in disordered potentials~\cite{Hulet2008}. Moreover, it has been used as evidence for superfluidity in optical lattices~\cite{Chin2006} and in rotating traps generating vortices~\cite{Madison2000, Zwierlein2005}.
	
	The relaxation of such excited states is of central importance to our understanding of the nexus of superfluid and macroscopic realms, such as superfluid helium flowing along rough surfaces. 
	For superfluid temperatures far below the transition point, it has been predicted that relaxation should occur free of dissipation as Kolmogorov-type turbulence~\cite{Svistunov95}.
	Experimentally, quantum dynamics of superfluids out of equilibrium were studied in various nonequilibrium realizations of superfluid helium ~\cite{Zhang2005,Bewley2008,Gomez2014}, but also in turbulent relaxation of driven, ultracold quantum gases~\cite{White2014,Navon2016}.
	However, different from interference of one- or two-dimensional quantum fluids, the role of long-range phase coherence in nonequilibrium quantum dynamics and hydrodynamics is challenging to access experimentally.
	Quenches, i.e., sudden changes of a system parameter, have proven to be a powerful tool for studying the nonequilibrium response of quantum systems.
	Examples include the collapse and revival of the matter-wave field of a BEC~\cite{Greiner2002}, the transport of atoms in optical lattices~\cite{Schneider2012, Meinert13}, or the response of quasi-particles upon a quench of interaction strength~\cite{Cetina2016}. 
	Beyond spatially homogeneous or periodic quenches, lattice systems have also been quenched into disorder, and the response was interpreted to show signatures of a Bose-glass phase~\cite{Meldgin2016}. 
	However, the behavior of long-range phase coherence following a quantum quench was not investigated in these works.
	
\begin{figure}
		\includegraphics[trim=10 15 25 0,clip]{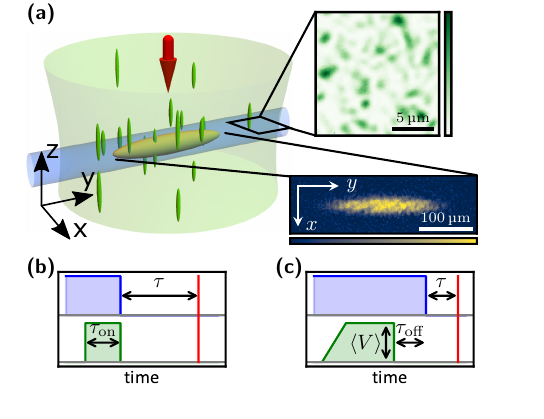}
		\caption{
			Schematic illustration of the experimental setup and measurement sequences.
			(a)~Experimental setup. The sample (yellow ellipsoid) is trapped in a superposition of an optical dipole trap (blue tube) and a magnetic saddle potential. The speckle beam (green volume) produces randomly distributed anisotropic grains. The insets show a section of the speckle intensity distribution in the $xy$-plane and an in-situ absorption image of a BEC in disorder.
			(b)~and (c)~Sequences for quenches into and out of disorder, respectively. Blue: optical dipole trap depth, green: disorder strength, red: imaging pulse. For measurements probing the expansion dynamics, the optical dipole trap and disorder potential are instantaneously extinguished and the gas is allowed to expand in the saddle potential for a variable time $\tau$ before the density distribution is recorded. The density dynamics are recorded in situ, i.e., with ${\tau=0}$.}
		\label{fig:figure1}
	\end{figure}

	Here, we study the response of a BEC to quenches of an optical disorder potential. Recording the in-situ density distribution and the expansion dynamics upon releasing the system from the trap, we can independently measure the responses of density and long-range phase coherence to the perturbation.
	Experimentally, we prepare quasi-pure molecular BECs of typically $4\times 10^5$ $^6$Li$_2$ molecules in an elongated harmonic trap (see Fig.~\ref{fig:figure1}) using standard techniques of laser and evaporative cooling~\cite{supps}. The trapping potential is a superposition of an optical dipole trap and a magnetic saddle potential, the latter being anti-confining in the $z$-direction. The trapping frequencies are $(\omega_x, \omega_y, \omega_z)={2\pi\times(164, 22.6, 107)\,}$Hz, leading to typical peak densities of $n_0=3.7 \times 10^{12}\,$cm$^{-3}$ at the cloud center. We tune the interaction utilizing a magnetic Feshbach resonance, enabling us to adjust the $s$-wave scattering length $a$ between the molecules~\cite{Grimm2007}. We use the gas parameter $n_0a^3$, which relates $a$ to the intermolecular distance ${\propto n_0^{-1/3}}$, to quantify the interaction strength. Subsequently, a repulsive optical speckle disorder potential $V(\bv r)$ composed of \SI{532}{\nano\meter} laser light and with a typical grain size ${\eta_{x,y}^2\times\eta_z=(\SI{750}{\nano\meter})^2 \times \SI{10}{\micro\meter}}$ is superimposed on the cloud, where $\eta_{x,y}$ and $\eta_z$ are the correlation lengths along the respective directions~\cite{Kuhn2007,supps}. We characterize the disorder strength by its spatial average $\avg{V}$, which also coincides with the standard deviation $\sqrt{ \avg{V^2}-\avg{V}^2}$ of the distribution.
	
	The dynamics of a  condensate in a speckle potential was first addressed experimentally~\cite{Lye:PRL2005} and theoretically~\cite{Modugno:PRA} for elongated samples.
	The introduction of the random potential affects the BEC in two ways. First, the density distribution $n$ readjusts to the altered external potential in order to minimize the energy of the system. Second, the phase is locally and dynamically shifted by ${\Delta \phi(\bv r) = V(\bv r) t / \hbar}$~\cite{pethick_smith_2008}, where $\hbar$ is the reduced Planck constant and $t$ the illumination duration. Importantly, for quantum fluids, both effects are coupled via the velocity field~\cite{pethick_smith_2008}
	\begin{equation}
		\bv v=\frac{\hbar}{m}\nabla\phi, \label{eq:velocity}
	\end{equation} 
	because a phase gradient is the source of a flow of density current $n \bv v$. The condensate can react to a spatial perturbation on a length scale given by the healing length ${\xi=1/\sqrt{8 \pi n_0 a}}$. In our experiment, the healing length at the trap center is below but of the order of the disorder grain size for all interaction strengths considered~\cite{supps}. Therefore, the condensate wave function resolves the spatial fluctuations of the speckle amplitude~\cite{SanchezPalencia2006}.\\

 	\section*{Density versus phase response}
\begin{figure*}[t]
		\includegraphics[trim=8 18 5 20,clip]{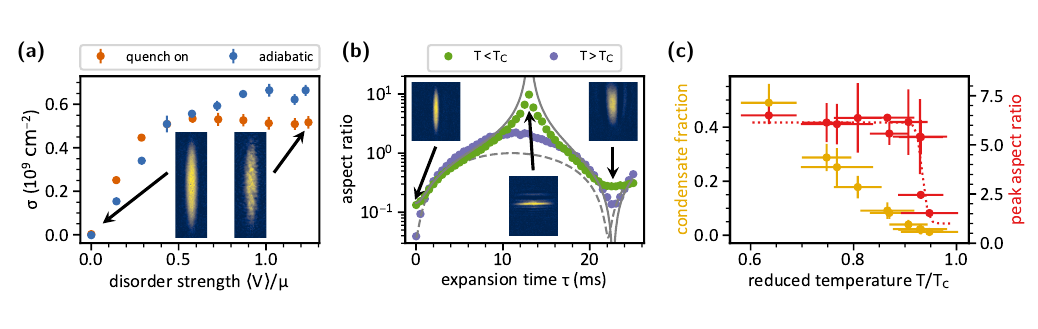}
		\caption{
			Observables used to quantify the density and long-range-phase-coherence response upon disorder quenches.
			(a)~Degree of density variation $\sigma$ in the long-time limit versus disorder strength, in units of the unperturbed chemical potential $\mu$ of the condensate, for ${n_0a^3=\num{1.1e-2}}$; the speckle is introduced both adiabatically (purple) and through a quench (green). Insets show absorption images for zero (left) and maximum (right) disorder strength.
			The error bars are standard deviations of \num{5} repetitions and different disorder realizations.
			(b)~Evolution of the cloud aspect ratio for $n_0a^3=\num{1.1e-2}$ and temperatures above and below $\tc$. The solid (dashed) line depicts the calculated trajectory for coherent hydrodynamic (ballistic) expansion~\cite{supps}. For short times, the measured trajectory for ${T<\tc}$ agrees well with the calculated one. For longer times, imaging aberrations due to the accelerating motion of the cloud along the imaging axis distort the measured aspect ratios, but qualitative agreement remains. Insets show absorption images for the case $T<\tc$ after \SI{0}{\milli\second}, \SI{13}{\milli\second}, and \SI{23}{\milli\second} expansion.
			(c)~Condensate fraction and peak aspect ratio versus reduced temperature for ${n_0a^3=\num{.4e-3}}$ in the absence of disorder.  Here, $\tc$ is the critical temperature of a noninteracting gas in a harmonic trap~\cite{pethick_smith_2008}. The dotted line serves as a guide to the eye.
		}
		\label{fig:figure2}
	\end{figure*}

	To unravel how density and long-range phase coherence relax under a disorder quench, we perform two different kinds of experiments. First, we measure the in-situ density distribution $n(x,y)$, column-integrated along the $z$-direction via resonant absorption imaging. Molecules are repelled from the regions of large potential, leading to spatial density variations, albeit no total fragmentation, the classical percolation threshold being far below the chemical potential~\cite{Pilati2010}.
	We then quantify the degree of density variations of these images as
	\begin{equation}
		\sigma = \sqrt{\langle \Delta n^2 \rangle - \langle \Delta n \rangle ^2},
	\end{equation}
	where ${\Delta n = n - n_\mathrm{fit}}$ is the difference between $n$ and a fitted 2D Thomas-Fermi profile $n_\mathrm{fit}$, and the brackets denote averaging over all pixels of the absorption image where ${n_\mathrm{fit}>0}$ (see Materials and Methods). In general, $\sigma$ is nonzero even in the absence of disorder, since the finite imaging resolution, as well as thermal effects, cause deviations of the density distribution from the expected Thomas-Fermi profile. In the following, we subtract this contribution and focus on the disorder-induced density response. The disorder effect on the density at long times is shown in Fig.~\ref{fig:figure2}~(a), where the speckle is either introduced adiabatically, within \SI{50}{\milli\second}, or through a quench. In the first case, the degree of density variations $\sigma$ increases monotonously with the disorder strength, while in the second case it grows faster for weak disorder but then saturates once the mean speckle potential $\avg{V}$ approaches half the chemical potential $\mu$. 
	Below, we use the time-evolution of $\sigma$ to quantify the response of the cloud's density to the disorder quench.

	Second, we investigate the long-range-phase-coherence response by studying the expansion of the quantum gas upon release from a confining potential.	
	The ensuing dynamics is entirely different from, e.g., a noninteracting, thermal cloud. The existence of a wave function $\psi$ implies collective dynamics similar to the hydrodynamic behavior of frictionless fluids~\cite{Giorgini1999,pethick_smith_2008}. Such coherent hydrodynamics leads to an inversion of the cloud aspect ratio during expansion from an anisotropic trap, which is a strong indication for BEC~\cite{Ketterle2008}. Coherent hydrodynamics originates from the existence of a macroscopic wave function, hence long-range coherence, and facilitates collective behavior such as quadrupole excitations. By contrast, the collisional hydrodynamic behavior in nondegenerate systems with strong interactions~\cite{Fletcher2018,Shvarchuck2003}, such as unitary gases, is caused by frequent scattering events during expansion and is therefore not connected to a macroscopic wave function. 
		
	Expansion is initiated by extinguishing the dipole trap beam and letting the cloud evolve in the stationary saddle potential. Coherent hydrodynamics manifests itself as a sharp peak in the aspect ratio during expansion~\cite{supps}, whose magnitude we use as a measure of long-range coherence, similar to a method proposed in~\cite{Shvarchuck2002}. Here, the aspect ratio is $R_x/R_y$, with $R_x$ and $R_y$ the Thomas-Fermi radii in $x$- and $y$-direction obtained from fits to the 1D integrated column-density distributions $n$. Fig.~\ref{fig:figure2}~(b) shows the dynamics of the aspect ratio for two cloud temperatures $T$ below and above $\tc$, the critical temperature for condensation. The aspect ratio of a quasi-pure BEC with ${T \ll \tc}$ features an initial exponential growth (due to the saddle-point confinement) followed by a pronounced peak with a value around ten at roughly a quarter trapping period along the long axis of the cloud. This behavior is attributed to the onset of quadrupole oscillations, indicating long-range coherence. This is emphasized by Fig.~\ref{fig:figure2}~(c), which directly connects the onset of coherent hydrodynamics, quantified by the peak aspect ratio during expansion, with the appearance of a condensate fraction in the cloud, and, therefore, of a macroscopic wave function.
	By contrast, the aspect ratio of a thermal cloud, for which ${T > \tc}$, varies slowly. The peak value of ${\approx \num{2.5}}$ is larger than one, which is the expected value for a gas with negligible interactions. We attribute this to a short initial phase of collisional hydrodynamics~\cite{Pedri2003} due to the relatively large $s$-wave scattering length of $a=\SI{2706}{\bohrradii}$, where \si{\bohrradii} is the Bohr radius.
	We will use the peak aspect ratio to quantify the breakdown and revival of long-range phase coherence in the system after the disorder quench.
	
	\begin{figure*}[t]
		\includegraphics[trim=27 15 25 15,clip]{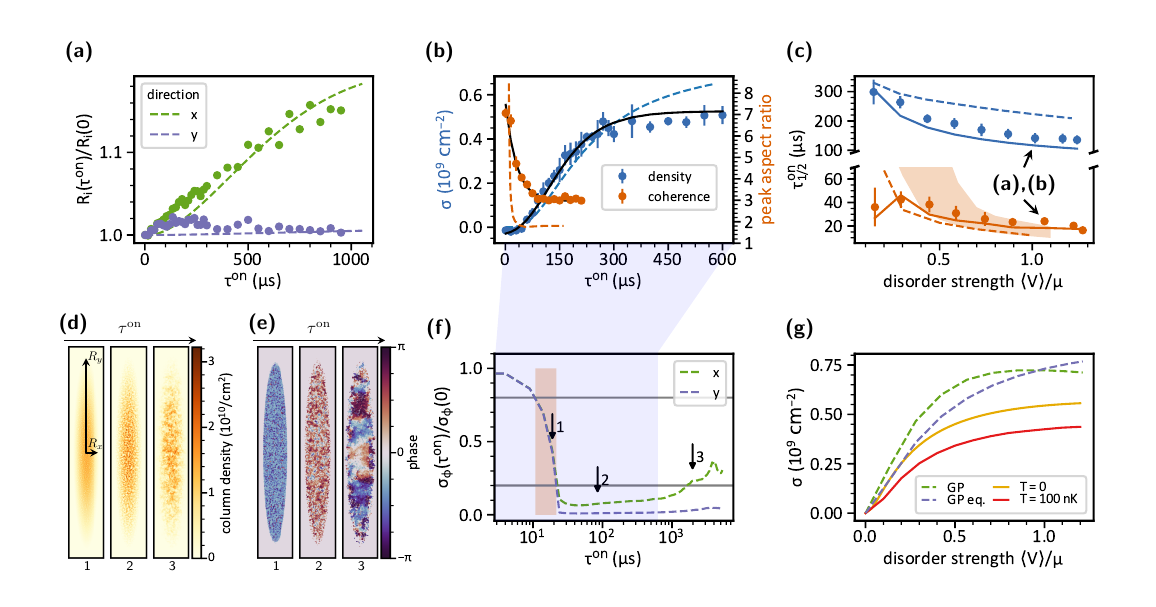}
		\caption{
			\textbf{Response of a quantum gas upon quenches into disorder.} Symbols refer to experimental data while dashed lines represent numerical results from the GP equation. (a) Time-resolved widths of the cloud, showing the broadening of the density distribution along the strongly confined axis. 
			(b) Density response (blue) and peak aspect ratio during expansion (orange) as a function of the exposure time $\tau^\mathrm{on}$, based on a time series of quantum gas images for ${\avg{V}\!/\mu = 1}$ and ${n_0a^3=\num{1.1e-2}}$. Solid black lines are fits to the data~\cite{supps}, dashed lines are results of numerical simulations. The error bars are standard deviations of five repetitions and different disorder realizations.
			(c)~Half-life periods of emerging density variations (blue) and the breakdown of coherent expansion (orange) for variable disorder strength and ${n_0a^3=\num{1.1e-2}}$, extracted from decay curves as in panel (b); the arrows point at the data points taken with the same experimental parameters as panels (a) and (b). Error bars denote fitting uncertainties. The solid lines depict $\trise$ (blue) and $\tbreakdown$ (orange) as defined in the text. $\tbreakdown$ incorporates the difference between the initial and final aspect ratio, which approaches zero for vanishing disorder strength~\cite{supps}. The orange shaded area displays the decay times of the phase-correlation length shown in panel~(f), where the gray lines indicate two different threshold values used to extract the decay time. The dashed lines show results of the numerical simulations.
			From the simulated wave-function dynamics, we extract (d) the integrated column density and (e) the phase in the central plane at different times. 
			(f) Normalized phase-correlation length, extracted from column-density and phase distributions.
			The arrows indicate the times $\tau^\mathrm{on}$ for which the column density and phase distributions are shown in~(d) and (e).
			(g) The computed density responses at equilibrium and in the steady state (dashed lines) compared with the predictions at equilibrium based on the local density approximation for a gas at equilibrium at ${T=0}$ and ${T=\SI{100}{\nano\kelvin}}$ (solid lines).
		}
		\label{fig:figure3}
	\end{figure*}
	
	We parallel these two experimental investigations by large-scale numerical simulations of the Gross-Pitaevskii (GP) equation~\cite{gross1961,pitaevskii1961} for a three-dimensional interacting quantum gas at zero temperature (see Materials and Methods for details). 
	The numerics takes into full account the specific properties of the speckle pattern used in the experiment.
	We emphasize that the study of the hydrodynamic expansion of the condensate is extremely challenging, because the initial wave function is affected by the disorder, implying that the analytical scaling ansatz~\cite{Kagan1997}  breaks down and the numerical solution of the GP equation in the saddle potential requires huge grids.
	From the numerical simulations, we extract time-resolved quantities after a disorder quench, namely the widths of the cloud and the column-integrated density, to directly compare with the experimental data. We obtain further insight into the many-body relaxation dynamics from the spatial and temporal dependence of the condensate phase. Although this quantity is not directly accessible in our experiment, the study of its autocorrelation function (see eq.~[12] in Materials and Methods) provides a natural explanation of the characteristic time scales for the loss and revival of long-range phase coherence observed in the experiment.

 	\section*{Response to quenches into disorder}
	First, we focus on the system's response upon quenches into disorder, tracing the decay of the unperturbed BEC properties. We instantaneously ($<\SI{1}{\micro\second}$) apply the speckle to a BEC for a time $\tau^\mathrm{on}$ (see Fig.~\ref{fig:figure1}~(b)). The density response is evaluated by imaging the cloud in-situ after $\tau^\mathrm{on}$ and recording the emerging density variations $\sigma(\tau^\mathrm{on})$. For the coherence response, the dipole trap is extinguished after $\tau^\mathrm{on}$, and we record the peak aspect ratio during expansion as a function of $\tau^\mathrm{on}$.
	
	Typical density- and coherence-response dynamics upon quenches into disorder are shown in Fig.~\ref{fig:figure3}.
	Globally, the cloud size (Fig.~\ref{fig:figure3}~(a)) exhibits an initial steady growth along the strongly confined $x$-axis, while the weakly confined $y$-axis is almost unaffected. This effect, which is well reproduced by our numerics, originates from the fact that atoms are pushed off by the repulsive speckle potential, and, due to the small Thomas-Fermi radius $R_{x}=\SI{18.5}{\micro\meter}$, they cannot rearrange along this axis without increasing the system size.

	By contrast, both phase and density variations respond much faster to the disorder quench (Fig.~\ref{fig:figure3}~(b)). Long-range coherence rapidly disappears with increasing illumination duration $\tau^\mathrm{on}$, while the density variations develop approximately one order of magnitude slower than the coherence responds. We find this behavior to prevail for all parameters studied here. 
	In the following, we denote the half-life period $\tau_{1/2}$ as the characteristic time after which density or coherence response have reached half their final value~\cite{supps}.
	In Fig.~\ref{fig:figure3}~(c), we summarize the half-life periods as a function of disorder strength. 
	We find that the half-life periods decrease with disorder strength.
	Besides, we have investigated the influence of interaction strength on the dynamics and found slightly larger response times for decreasing interaction strength (see \cite{supps}).

	An intuitive picture of the underlying mechanisms can be obtained from simple energy arguments (see Materials and Methods for details).
	For the density, after switching on the speckle, the random potential causes a spatially varying accumulation of phase and, therefore, a local velocity field according to Fig.~\ref{eq:velocity}. We are interested in the typical time $\trise$ after which the flow has traversed a given distance, which we set to the resolution of our imaging system ${\alpha=\SI{2.2}{\micro\meter}}$. 
	Thus, we estimate the mean velocity from the average gradient ${\propto \avg{V}\!/\eta_{x,y}}$ of the local disorder potential, yielding ${\trise \propto \sqrt{\alpha\eta_{x,y}/\avg{V}}}$. This time-scale is indicated in Fig.~\ref{fig:figure3}~(c) as solid blue line.
	Furthermore, we attribute the breakdown of coherent hydrodynamics to the phase imprint onto the BEC by the disorder potential, which is given by ${\Delta\phi(\bv r)=V(\bv r)\tau^\mathrm{on}/\hbar}$. The phase pattern changes on length scales of the disorder correlation length, which is much smaller than the size of the quantum gas, and roughly a factor of two larger than the healing length of the condensate. Thus, the quench initiates a rapid and fine-grained phase evolution, eventually leading to dephasing between different locations within the cloud. From the mean phase difference ${\avg{\delta \phi} = \avg{V} \tau^\mathrm{on}/\hbar}$ between two points in the BEC, we deduce the time scale for breakdown of coherent hydrodynamics ${\tbreakdown\propto\hbar/\avg{V}}$. This time scale is indicated in Fig.~\ref{fig:figure3}~(c) as a solid orange line and reproduces well the trend of the experimental data.
			
	\begin{figure*}[t]
		\includegraphics[trim=10 15 15 15,clip]{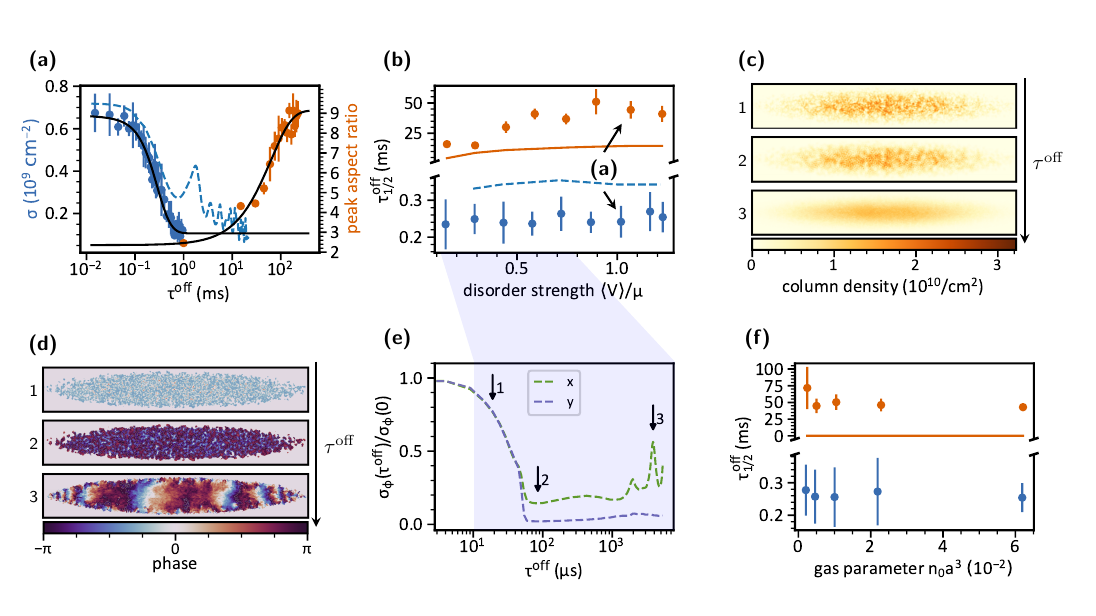}
		\caption{
			\textbf{Response of a quantum gas upon quenches out of disorder.} Symbols refer to experimental data while dashed lines represent numerical results from the GP equation.
			(a)~Quantum gas dynamics showing decaying density variations (blue) and revival of coherent hydrodynamic expansion (orange) for ${\avg{V}\!/\mu = 1}$ and ${n_0a^3=\num{1.1e-2}}$. The error bars are standard deviations of five repetitions and different disorder realizations. The dashed line is obtained from the numerical model and black lines are fits to the experimental data~\cite{supps}.
			(b) Half-life periods of such dynamics as a function of disorder strength for a gas density of ${n_0a^3=\num{1.1e-2}}$, where the error bars denote fitting uncertainties. For comparison, the solid line depicts $\treoccurrence$ as defined in the Materials and Methods, indicating the time scale to cross the long axis of the BEC with the speed of sound. Clearly, the time scale for revival of quantum hydrodynamics is much longer than the time scale $\treoccurrence$. Here, $\treoccurrence$ incorporates the difference between the initial and final aspect ratio. 
			(c) and (d) show three different snapshots of the numerically simulated column density and in-plane phase distribution, respectively, following the quench.
			(e) Normalized phase-correlation length $\sigma_\phi$ extracted from the two time-resolved distributions (c) and (d), showing the absence of long-range coherence within the simulated times. The arrows in (e) indicate the times $\tau^\mathrm{off}$ for which the column density and phase distributions are shown in~(c) and (d).
			(f) displays the half-life periods for variable interaction strength and ${\avg{V}\!/\boltzmann=\SI{145}{\nano\kelvin}}$, where the solid line again indicates the time scale $\treoccurrence$ for comparison.	
		}
		\label{fig:figure4}
	\end{figure*}	

	Next, we compare our GP numerics with the experimental data. We obtain time-resolved column density and in-plane phase distributions,
	as shown in Fig.~\ref{fig:figure3}~(d) and~(e), respectively. 
	The density response is then computed by first convolving the density data with a Gaussian function of width \SI{2.2}{\micro\meter}, to account for the limited resolution of the imaging system (see \cite{supps}). The result agrees reasonably well with the measured data, although numerics predicts 
	a larger value of $\sigma$ in the steady state, which in turn leads to a larger half-life period, as shown in Fig.~\ref{fig:figure3}~(b) and (c). 
	In Fig.~\ref{fig:figure3}~(g) we plot the density response as a function of the disorder strength, calculated both at equilibrium and in the steady state, following the disorder quench. The obtained results are in good agreement with the experimental data shown in Fig.~\ref{fig:figure2}~(a).
	A closer comparison reveals that the numerics overestimates $\sigma$, especially when the disorder is strong. 
	We attribute this residual difference to finite temperature effects, which are neglected in the GP equation. To understand the role of temperature, we compute the density response of the disordered gas at equilibrium via the local-density approximation (LDA), both at ${T=0}$ and at ${T=\SI{100}{\nano\kelvin}}$, by using the Hartree-Fock approach (see Materials and Methods). The results are plotted in Fig.~\ref{fig:figure3}~(g) with solid lines. Two comments are in order here. 
	First, while for weak disorder the LDA result for the disorder-induced density response at zero temperature is indistinguishable from the prediction of the GP equation, for strong disorder the LDA result falls below it.
	Indeed, LDA underestimates the atom density in the center of the trap, which contributes mostly to the signal, while it overestimates the density at the periphery. Second, even if the gas is in the quantum degenerate regime, thermal effects can broaden the density distribution and deplete the density response.\\
	Let us now discuss the GP results for the hydrodynamic expansion of the BEC, after exposure to the speckle potential. The calculated peak aspect ratio and the associated half-life period, displayed in Fig.~\ref{fig:figure3}~(b) and (c) respectively, reproduce the experimental trend, although with much faster decay of long-range phase coherence for short illumination times, because the GP numerics does not account for imaging aberration effects. 
	Notice that the difference in the peak aspect ratio observed for large $\tau^\mathrm{on}$ is mainly due to the large in-plane grid spacings used for the expansion dynamics~\cite{supps}.  Notwithstanding, the separation of time scales for density and phase relaxations is recovered, and even accentuated, by the GP numerics.
	
	The disorder-induced loss of long-range coherence can be directly related to the scrambling of the condensate phase before the expansion. From the snapshots of 
	density and phase distributions as in Fig.~\ref{fig:figure3}~(d) and~(e), we extract the time-resolved phase correlation function~\cite{supps} of the BEC. 
	The result is shown in Fig.~\ref{fig:figure3}~(f). We see that the phase-correlation length $\sigma_\phi$ drops around ${\tau^\mathrm{on}=\SI{20}{\micro\second}}$, which roughly 
	corresponds to the observed half-life period in Fig.~\ref{fig:figure3}~(c). 
	We repeat the analysis for different disorder strengths. We consider a range of phase-correlation lengths between $0.2 \ldots 0.8$ of the maximum value and extract the time scale on which these correlation lengths are reached, indicated by an orange shaded area in Fig.~\ref{fig:figure3}~(c). The numerical values yield indeed the correct order of magnitude for the time response.
	We conclude that the density and phase responses of the quantum gas to disorder quenches can be unraveled by the two measurement methods. 
	Overall, the loss of long-range phase coherence is typically one order of magnitude faster than the density relaxation.

 	\section*{Response to quenches out of disorder}
		Next, we consider the case when the quantum system relaxes after release from an initially disordered state and ask the question when an unperturbed density distribution and long-range coherence are reestablished.
		Quenches out of disorder are realized by slowly introducing the speckle during a \SI{50}{\milli\second} linear ramp, in order to minimize excitations in the gas, and subsequently waiting for \SI{100}{\milli\second} to let it equilibrate. Then we suddenly extinguish the speckle and wait for a variable time $\tau^\mathrm{off}$, during which the system can relax (Fig.~\ref{fig:figure1}~(c)), before probing the density variations or expansion dynamics, respectively.
	
		For the density response, we do not find any dependence of the half-life period ${\tau_{1/2}^\mathrm{off}\approx\SI{250}{\micro\second}}$ on either disorder or interaction strength (see Fig.~\ref{fig:figure4}~(b) and (f)).
		The final value of $\sigma$ after the longest measured wait time of ${\tau^\mathrm{off}=\SI{1}{\milli\second}}$ is up to $0.1\times 10^{9}$cm$^{-2}$ above its value in the clean case for large disorder strengths (see Fig.~\ref{fig:figure4}~(a)).
	
		By contrast, we find that it takes two orders of magnitude longer to restore long-range phase coherence, with a peak aspect ratio comparable in magnitude to the disorder-free case. 
		This is consistently observed for all disorder strengths applied, as shown in Fig.~\ref{fig:figure4}~(b). 
		Here, too, simple arguments allow relating the observed time scales to the energy scales of the system.
		The long time to reestablish coherent hydrodynamics can be compared to the longest time scale in the system, i.e., the time $\treoccurrence$ a signal needs to traverse the long axis of the cloud with the speed of sound, ${\treoccurrence=2 R_y / v_\mathrm{s}}$, where $R_y$ is the largest Thomas-Fermi radius of the BEC and ${v_\mathrm{s}=\sqrt{\mu/m}}$ is the maximum speed of sound at the center of the cloud. Furthermore, we observe that both the density and the coherent response are rather independent of interactions in the gas, see Fig.~\ref{fig:figure4}~(f).
		Importantly, there are no significant particle losses during $\tau_\mathrm{off}$ (see \cite{supps}). This excludes speckle-induced heating and subsequent evaporation as the origin of the breakdown and revival of coherent hydrodynamics.
		In the transient regime after the quench, the widths of the BEC along the $x$- and $y$-directions remain  unchanged. The numerics shows that the BEC shrinks along the $z$-direction, where the disorder potential varies slowly.
		
		The  GP simulations predict a rapid decay for the density response (Fig.~\ref{fig:figure4}~(a)), which ultimately saturates to a nonzero value, as also found in the experiment. For sufficiently strong disorder $(\langle V\rangle/\mu \gtrsim 0.5)$, the calculated density response features large amplitude oscillations in the long time limit (after \SI{1}{\milli\second}), which are absent in the experimental data. The origin of this effect can be traced back to a corresponding oscillation of the Thomas-Fermi radii, especially along the strongly confined $x$-axis. The compression of the cloud leads to an enhancement of the peak atom density, which in turn causes an increase in the density response. In the experiment, this collective mode is probably damped by the coupling between the BEC and the thermal component of the cloud~\cite{fedichevDamping1998}. For this reason, the calculated half-life period for the density response (Fig.~\ref{fig:figure4}~(b)) is slightly larger than in the experiment.
		On the other hand, the calculation of the peak aspect ratio for values of $\tau^\textrm{off}$ of the order of tenths or hundreds of milliseconds is numerically heavy, due to growing errors, which affect the subsequent expansion dynamics. For this reason, we leave it for future studies. 
		In (Fig.~\ref{fig:figure4}~(c) and (d)) we display some snapshots of the column density and phase distributions following the disorder quench, from which we extract the phase correlation function (Fig.~\ref{fig:figure4}~(e). 
		Within the simulated time intervals, we observe a progressive locking of the local condensate phase along the $x$-axis.  
		By contrast,  the phase varies rapidly along the weakly confined $y$-direction, signaling the presence of phonon-like excitations. 

 	\section*{Conclusions}
		We have performed an experimental and numerical study of the far-from equilibrium dynamics of a molecular BEC subject to a quench of the disorder potential. We have found that density and long-range phase coherence respond to the perturbation on different time scales.
		Specifically, the long times needed to restore long-range phase coherence might indicate the decay of a complex phase pattern toward an ordered phase, where, for instance, phase boundaries or vortices originating from the disorder quench are topologically robust and need a relatively long time to decay.
		
		This picture directly connects our observation to the recently reported absence of hydrodynamic behavior in BECs, where turbulence was introduced by applying a spatially homogeneous, oscillating force~\cite{Navon2016, Henn2009}. 
		Numerical simulations show that random phase imprints, spatially varying on a length scale slightly larger than the healing length, also result in turbulent flow~\cite{Kobayashi2005}. Turbulence and accompanying vortices can be rather persistent with lifetimes exceeding several \SI{100}{\milli\second}~\cite{Zwierlein2005, Navon2016}. This suggests that the phase dynamics ensuing after a disorder quench might generate turbulent flow that takes a relatively long time to decay before long-range phase coherence is established. The fact that we do not see a sign of this in the density distributions for times longer than ${\sim\SI{1}{\milli\second}}$ after quenches might be explained by the limited optical resolution of ${\alpha=\SI{2.2}{\micro\meter}}$ of our imaging system and the integration of the density along the $z$-axis. In turbulent flow, an energy cascade~\cite{Navon2016} could transfer excitations to smaller length scales we cannot resolve.
		
		In the future, it will be interesting to study the dynamical response of quantum gases along the BEC-BCS crossover to explore the impact of quenched disorder on resonantly interacting superfluids. Our system is also ideally suited to follow further the phase dynamics and its dependence on quench parameters. Finally, our work calls for more  refined numerical simulations of the dynamics of the disordered BEC including finite temperature effects.
	
	\section*{Acknowledgments}
		We thank B.\ G\"anger and J.\ Phieler for help in the construction of the apparatus, and J. Anglin for valuable discussions.  We also acknowledge comments by M.~Modugno and S.~Pilati on the manuscript, which helped us improving the presentation. 
		This work was supported by the Deutsche Forschungsgemeinschaft (DFG, German Research Foundation) via the Collaborative Research Center SFB/TR185 (Project No.\ 277625399). B.N.\ received support from a DFG Fellowship through the Excellence Initiative by the Graduate School Materials Science in Mainz (GSC 266). J.K.\ was supported by the Max Planck Graduate Center with the Johannes Gutenberg-Universität Mainz (MPGC). This work was granted access to the HPC resources of CINES under allocations 2019-A0060507629 and 2020-A0080507629 supplied by GENCI (Grand Equipement National de Calcul Intensif).
	
	\newpage
	
	\appendix
	
	In the following, details on the experimental procedure and the theoretical models, as well as additional data are given.

	\subsection*{Setup and sequence}
		A general overview of our experimental apparatus is presented in~\cite{Gaenger2018}. We prepare quantum gases in the BEC-BCS crossover regime by forced evaporative cooling of fermionic $^6$Li atoms in an equal mixture of the two lowest-lying Zeeman substates of the electronic ground state $^2\mathrm{S}_{1/2}$. Evaporation takes place in a hybrid magnetic-optical trap at a magnetic field of $\SI{763.6}{\gauss}$ on the repulsive side of a Feshbach resonance centered at \SI{832.2}{\gauss}~\cite{Zuern2013}, where atoms of opposite spin form bosonic molecules that eventually condense into a BEC. After evaporation, the sample is held at constant trap depth for \SI{250}{\milli\second} to ensure thermal equilibrium before the magnetic field is linearly ramped to its final value during \SI{200}{\milli\second}. We employ resonant high-intensity absorption imaging~\cite{Reinaudi2007} to extract the column density distribution in the $x$-$y$ plane. From bimodal fits to the in-situ density distribution~\cite{Naraschewski1998} at \SI{680}{\gauss}, we are not able to discern a thermal fraction.

		The hybrid trap consists of an optical dipole trap and a magnetic saddle potential, which provides weak (anti-) confinement in ($z$-) $x$- and $y$-direction, whereas the optical trap strongly constrains the cloud along $x$ and $z$. Since the saddle potential is an accessory to the magnetic field used to address the Feshbach resonance, its curvature depends on the field magnitude. For all experiments presented here, the combined trapping frequencies of the optical and magnetic trap are ${\omega_x=2\pi\times\SI{164}{\hertz}}$ and ${\omega_z=2\pi\times\SI{107}{\hertz}}$. $\omega_y$ is listed in Tab.~S.1 for the different magnetic fields addressed.

		The speckle potential is created by passing a laser beam of wavelength \SI{532}{\nano\meter} through a diffusive plate and focusing the light, using an objective with numerical aperture \num{0.29}, onto the atoms. They experience a repulsive and spatially random (but temporally constant) dipole potential $V$, which we characterize by its average $\avg{V}$ at the focal point of the objective. The typical grain size of the speckle is given by the Gaussian-shaped autocorrelation function of the potential with $1/e$ widths (correlation lengths) ${\eta_{x,y}=\SI{750}{\nano\meter}}$ transversely to and ${\eta_z=\SI{10}{\micro\meter}}$ along the beam propagation direction. As the speckle beam has a Gaussian envelope with waist \SI{850}{\micro\meter}, the disorder potential is slightly inhomogeneous with less than \SI{5}{\percent} variation of $\avg{V}$ across the typical cloud size. We change the specific disorder realization by slightly rotating the speckle pattern as a whole between repetitions. For that reason, the diffusive plate is attached to a motorized rotation mount. This allows us to measure disorder-averaged quantities that are independent of the microscopic details of any specific disorder realization. 
		Switching on and off the speckle potential and letting the BEC equilibrate subsequently, we do not find a measurable increase of temperature.

	\subsection*{Measurement of density variation}
		\begin{figure}
			\includegraphics[trim=50 0 30 0,clip]{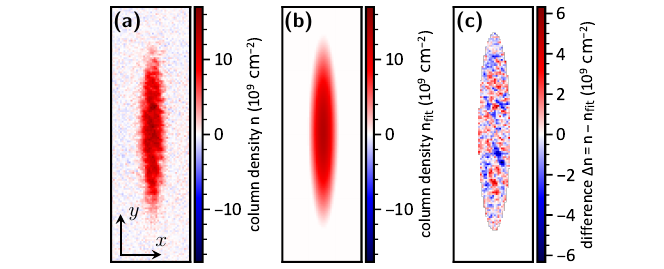}
			\caption{
				Calculation of density variation $\sigma$ from exemplary density profile obtained at interaction strength $n_0a^3=\num{1.1e-2}$ and disorder strength $\avg{V}\!/\mu=1.2$.
				(a)~Measured density distribution $n$.
				(b)~Fitted Thomas-Fermi profile $n_\mathrm{fit}$.
				(c)~Difference $\Delta n=n-n_\mathrm{fit}$ in the region where $n_\mathrm{fit}>0$.}
			\label{fig:supp_figure0}
		\end{figure}
	
		We quantify the degree of density variation of a measured column density distribution $n$ as $\sigma={\sqrt{\langle \Delta n^2 \rangle - \langle \Delta n \rangle ^2}}$ with ${\Delta n = n - n_\mathrm{fit}}$. $n_\mathrm{fit}$ is a smooth, two-dimensional Thomas-Fermi profile
		\begin{equation}
		n_\mathrm{fit} \propto
			\begin{cases}
				p^{3/2} & p > 0 \\
				0 & \, \text{else},
			\end{cases}
		\end{equation}
		with ${p=1-(x/R_x)^2-(y/R_y)^2}$, fitted to $n$. The brackets denote averaging over all pixels with $n_\mathrm{fit}>0$. Due to imaging aberrations and inhomogeneities of the imaging setup, $\sigma$ is larger than zero even for density profiles without disorder. We correct for that by subtracting this offset.
		
		To extract the half-life period from the density response dynamics (see Fig.~\ref{fig:figure2}~(a)), we fit the time series with a Gompertz function ${\propto \exp(-b \exp(-ct))}$~\cite{weissteingompertz}. The half-life period is obtained by calculating ${\tau_{1/2}=-\log(\log(2)/b)/c}$, where $\log$ is the natural logarithm.
		
	\subsection*{Cloud expansion into a saddle potential}
		\begin{figure}
			\includegraphics{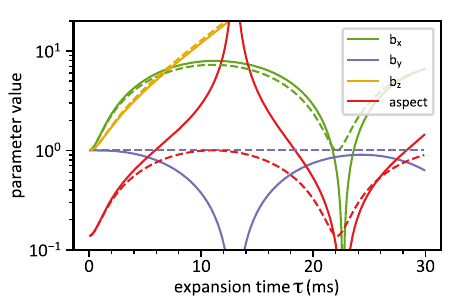}
			\caption{ Scaling parameters and aspect ratio for coherent hydrodynamic (solid lines) and ballistic (dashed lines) expansion into the saddle potential for a magnetic field of \SI{763.6}{\gauss} in our setup.}
			\label{fig:supp_figure1}
		\end{figure}
	
		The time evolution of a BEC with initial density distribution ${n(\bv r, t=0)}$ in a harmonic trap with time-dependent frequencies $\omega_i(t)$ (${i=x,y,z}$) can be described in terms of a scaling transform ${n(\bv r, t) = n(x/b_x, y/b_y, z/b_z, t)/b_x b_y b_z}$~\cite{Kagan1997}. 
		The scaling parameters  $b_i(t)$ are obtained from the solution of	
		\begin{equation}
			\label{eq:hydrodynamic_eq}
			\ddot{b}_i = -\omega_i^2(t) b_i + \frac{\omega_i(0)^2}{b_i b_x b_y b_z}
		\end{equation}
		with boundary conditions ${b_i(0)=1}$ and ${\dot{b}_i(0)=0}$. For our system, 
		$(\omega_x(0), \omega_y(0), \omega_z(0))=2\pi\times(\textrm{164, 22.6, 107})$Hz for \SI{763.6}{\gauss}.
		With decreasing magnetic field, also $\omega_y(0)$ decreases slightly (see Tab.~S.1) while $\omega_x(0)$ and $\omega_z(0)$ are solely determined by the optical trap. Upon extinction of the dipole trap at ${t=0}$, the trapping frequencies instantaneously take on the values $\omega_{x,y,z}(t)=\omega_y(0), \omega_y(0), \iu\sqrt{2}\omega_y(0)$. The imaginary frequency reflects the anti-confining nature of the saddle potential along $z$. Note that (\ref{eq:hydrodynamic_eq}) neglects the contribution of the quantum pressure ${\propto\nabla^2\sqrt{n}}$~\cite{pethick_smith_2008}. Fig.~\ref{fig:supp_figure1} shows the dynamics of the scaling parameters during expansion. 
		The  reduction of the trapping frequency along the $x$ axis causes an initial rapid expansion of the cloud in this direction and a concomitant contraction along the y-axis,   leading to the  inversion of the aspect ratio. This effect is further amplified by  the presence of the saddle potential along the $z$ axis, stretching the cloud ever-increasingly.
		In contrast, a noninteracting cloud does not exhibit collective behavior and each particle escapes with its momentary velocity at the time of release. This facilitates an analytical description of such ballistic expansion dynamics in terms of a scaling transform~\cite{Ketterle2008}, the corresponding trajectories are displayed in Fig.~\ref{fig:supp_figure1}. The most distinct feature of hydrodynamic expansion, as compared to ballistic expansion, is the contraction and subsequent expansion of the cloud along its initially longer axis,  causing an inversion of the aspect ratio.

	\subsection*{BEC properties}
		In the absence of disorder, the BEC is well described by the Thomas-Fermi approximation, since the condition ${Na/\bar{a} \gg 1}$~\cite{pethick_smith_2008} is fulfilled for all magnetic fields. Here, $N$ is the number of molecules and ${\bar{a}=\sqrt{\hbar/m\bar{\omega}}}$ the harmonic oscillator length that corresponds to the geometric mean of the trapping frequencies $\bar{\omega}$. The cloud radii are given by ${R_i=(2\mu/(m\omega_i^2))^{1/2}}$, where ${\mu=(15Na/\bar{a})^{2/5}\hbar\bar{\omega}/2}$ is the chemical potential. The speed of sound and the healing length at the trap center are given by ${v_\mathrm{s}=\sqrt{\mu/m}}$ and ${\xi=1/\sqrt{8\pi n_0 a}}$, respectively.     The typical numbers for our experiment are given in Tab.~\ref{tab:parameters}. The comparison with the numerical simulation discussed in the main text concerns the configuration with the largest scattering length, corresponding to the last column of Tab.~\ref{tab:parameters}.
		
		\begin{table}[h]
			\begin{center}
				\begin{tabular}{l | r | r | r | r | r}
					magnetic field (G) & 680.0 & 700.0 & 720.0 & 740.0 & 763.6 \\
					\hline
					$\omega_y/2\pi$ (\si{\hertz}) & 21.4 & 21.7 & 22.0 & 22.3 & 22.6 \\
					$a$ ($a_0$) & 743 & 982 & 1310 & 1784 & 2706 \\ 
					$n_0a^3$ ($10^{-3}$) & 0.4 & 0.9 & 1.8 & 3.9 & 11.0 \\
					$N$ ($10^5$) & 3.0 & 3.3 & 3.7 & 3.9 & 4.3 \\
					$Na/\bar{a}$ ($10^3$) & 3.5 & 5.2 & 7.6 & 11.1 & 18.4 \\
					$n_0$ ($10^{12}\si{\per\centi\meter\cubed}$) & 7.0 & 6.2 & 5.4 & 4.6 & 3.7 \\
					$\mu / \boltzmann$ (\si{\nano\kelvin}) & 132 & 156 & 183 & 214 & 263 \\
					$\xi$ (\si{\nano\meter}) & 380 & 350 & 325 & 300 & 270 \\
				\end{tabular}
			\end{center}
			\caption{Overview of parameters for different magnetic fields. Scattering lengths are taken from~\cite{Zuern2013}.}
			\label{tab:parameters}
		\end{table}

	\subsection*{Numerical simulations of the experiment}
		\label{sec:num}
		\textbf{Modeling the disorder:}
		The disorder potential used in the experiment is a single, blue-detuned, anisotropic speckle with an on-site probability distribution given by the Rayleigh law, 
		$P(V)=e^{-V/\langle V \rangle} \Theta(V)/\langle V \rangle$,
		where $\Theta(x)$ is the Heaviside function. 
		The speckle pattern is generated by a laser beam parallel to the $z$-axis and is characterized by the spatial correlation function~\cite{Goodman:07}
		\begin{equation}\label{spec0}
			C(\mathbf r)=\frac{\langle V(\mathbf r) V(0)\rangle}{\langle V^2(0)\rangle}= \left|\frac{f(\mathbf r)}{f(0)}\right|^2,
		\end{equation}
		where
		\begin{equation}\label{spec1}
			f(\mathbf r)=\int_0^\pi e^{i 2 \pi \frac{z}{\lambda_L} \cos \theta }J_0\left(2 \pi \frac{\rho}{\lambda_L} \sin \theta \right)h(\theta)\sin \theta \mathrm{d}\theta.
		\end{equation}
		Here, $\lambda_L=\SI{532}{\nano\meter}$ is the laser wavelength, $J_0(x)$ is the zero-order Bessel function and $\rho=\sqrt{x^2+y^2}$. The function $h(\theta)$ in eq.~(\ref{spec1}) depends on the specific experimental setup. Following Ref.~\cite{Volchkov:PRL2018}, we write it as 
		\begin{equation}\label{spec2}
			h(\theta)=\exp{\left(-\frac{2\tan^2 \theta} {\theta_0^2}\right)} \Theta(\tan \theta_\textrm{max}-|\tan \theta|),
		\end{equation}
		where $\theta_\textrm{max}$ defines the maximal numerical aperture $NA=\sin(\theta_\textrm{max})=0.29$
		of the objective, while $\theta_0$ is a free parameter.
		Its specific value is fixed to reproduce the measured correlation length $\eta_{x,y}=\SI{750}{\nano\meter}$ of the speckle in the $xy$-plane, which is defined as $C(x=\eta_{x,y},0,0)=1/e$. By substituting eq.–(\ref{spec2}) into (\ref{spec1}) and carrying out the numerical integration over the angle, we find that this condition is satisfied for $\theta_0= 0.25$. 
		
		The speckle potential has been generated numerically on a 3D grid with uniform spacings $\Delta x=\Delta y=\Delta z=0.3 \lambda_L$. 
		This is done  by first generating at each site $j$ of the grid a complex field $E_j$, whose real and imaginary parts are uncorrelated normally distributed random variables with zero mean and unity variance. The total electric field at position $\mathbf r$ is obtained by convolving the random field with an appropriate mask $\mathcal P(\mathbf k)$ in Fourier space~\cite{Pilati2010}:
		\begin{equation}
			E(\mathbf r)=\sum_\mathbf k E(\mathbf k) \mathcal P(\mathbf k) e^{i \mathbf k \cdot \mathbf r},
		\end{equation}
		so that the resulting speckle potential $V(\mathbf r)=\avg{V}|E(\mathbf r)|^2/\langle |E|^2\rangle$ obeys the correlation function in Eqs. (\ref{spec0}-\ref{spec2}). This yields $\mathcal P(\mathbf k)=\delta(|\mathbf k|-2\pi/\lambda_L) \sqrt{h(\theta)}$, where the angle $\theta$ is defined by $k_z=|\mathbf k| \cos\theta$.
		
		\vspace{0.2cm}
		\textbf{Numerical solution of the Gross-Pitaevskii (GP) equation:} At zero temperature, the BEC is described by a macroscopic wave function $\psi(\mathbf r,t)$ obeying the non-linear mean-field GP equation~\cite{gross1961,pitaevskii1961}
		\begin{equation}\label{GP}
			i \frac{\partial \psi}{\partial t}=\left[-\frac{\hbar^2}{2m} \nabla^2 +V_\textrm{tr}(\mathbf r) + V(\mathbf r)+g |\psi(\mathbf r,t)|^2\right]\psi(\mathbf r,t),
		\end{equation} 
		where $V_\textrm{tr}(\mathbf r)=m(\omega_x^2 x^2+\omega_y y^2+\omega_z^2 z^2)/2 $ is the harmonic confining potential and $g=4\pi \hbar^2 a/m>0$ is the strength of the 
		boson-boson repulsion. The amplitude of the wave function is normalized according to $\int |\psi(\mathbf r,t)|^2 \mathrm{d}^3r=1$ and is related to the 3D particle density through $n_{3D}(\mathbf r,t)=N |\psi(\mathbf r,t)|^2$.
		The ground state density profile of the BEC is obtained from the GP equation via imaginary time propagation.
		In the absence of disorder, the Thomas-Fermi radii of the condensate are $R_{x}=\SI{18.5}{\micro\meter}$, $R_{y}=\SI{134.4}{\micro\meter}$ and $R_{z}=\SI{28.4}{\micro\meter}$. 
		The numerical integration of the GP equation for the trapped gas is carried out on a grid of fixed dimensions $(N_x, N_y, N_z)=(300, 2200, 450)$
		and uniform spacing $\Delta x=\Delta y=\Delta z=\SI{0.16}{\micro\meter}$. We use the open-source code presented in Refs.~\cite{YOUNGS2017503,MURUGANANDAM20091888}, implementing
		the split-step Crank-Nicholson method.  Simulating one millisecond of time evolution for a given disorder configuration requires between one and two hours of wall-clock time on an Intel Haswell node with 24 CPU cores.\\
		The maximum value of the aspect ratio during the expansion is reached after (roughly)  $t=\SI{13}{\milli\second}$. GP simulations for such
		long time scales are difficult, also because 
		 the BEC widths along the $x$ and $z$ axes grow rapidly with time, requiring exceedingly large grids. In particular the anti-confining potential accelerates the particles outwards, implying that a fine mesh is necessary to follow the rapid spatial oscillations of the wave function and avoid numerical instability at the condensate tails. We simulate the expansion dynamics by increasing the lattice spacing along the transverse directions, $\Delta x=\Delta y=\SI{0.8}{\micro\meter}$, while keeping unchanged the spacing along the $z$-axis.
		We further speed up the calculation by adapting dynamically the grid shape to match the density profile, reaching sizes up to $(N_x, N_y, N_z)=(520, 520, 7800)$ without producing memory errors. Each simulation of the expansion dynamics takes between 16 and 24 hours of wall-clock time, depending on the specific value of the exposure time $\tau^\textrm{on}$.			
		The use of a coarser grid introduces some approximation, in particular the peak aspect ratio for large exposure time $\tau^\mathrm{on}$ is underestimated, resulting in an apparent discrepancy with the experimental data. 
		
		In order to extract the characteristic time $\tau^\mathrm{on}_{1/2}$ from the simulated expansion dynamics, we calculate the time $\tau^\mathrm{on}$ after which the peak aspect ratio during expansion has dropped to half its initial value. To this end, the values of the aspect ratio for varying $\tau^\mathrm{on}$ are interpolated and we only consider values of the aspect ratio below 10, since larger values are experimentally not accessible due to imaging aberrations.
		
		\vspace{0.2cm}
		\textbf{Finite temperature effects:} 
		The GP equation assumes that the gas is at zero temperature. To understand how temperature affects the density response of the gas, we focus on the
		equilibrium case, where the speckle is loaded adiabatically. 
		Assuming that the condensate healing length is small compared to the speckle grain size in the $x$- and $y$-directions, we can treat the disorder as a slowly varying potential and use a finite-temperature generalization of the local-density approximation (LDA), based on the Hartree-Fock approach. Within this theory, the condensate and the thermal densities, indicated respectively by $n_c$ and $n_T$, satisfy the coupled equations~\cite{Pilati2010}
		\begin{align}\label{eq:systemeq}
			n_c(\mathbf r)&=\frac{1}{g} (\mu-V_\mathrm{ext}(\mathbf r)-2 gn_T(\mathbf r)),\\
			n_T(\mathbf r)&=\int \frac{\mathrm{d}^3p}{(2\pi\hbar)^3} \frac{1}{e^{\beta (\frac{p^2}{2m}-\mu +V_\mathrm{ext}(\mathbf r)+2 g (n_c(\mathbf r)+ n_T(\mathbf r))}-1},\nonumber
		\end{align}
		where $V_\mathrm{ext}=V_\textrm{tr}+V$ is the total external potential acting on the atoms and $\beta=1/(k_B T)$, with $k_B$ being the Boltzmann constant. The chemical potential $\mu$ in the above equations must be fixed from the normalization condition $N=\int \mathrm{d}^3r n_{3D}(\mathbf r)$, where $n_{3D}(\mathbf r)=n_c(\mathbf r) +n_T(\mathbf r)$ is the total density. From the latter, we extract the column density and the density response, following the same procedure  used for the  GP numerics.

	\subsection*{Description of timescales}
		We extract the half-life periods of the density (coherence) response by fitting the time series with Gompertz (exponential) functions~\cite{weissteingompertz}, which we have found to adequately describe all data.
		
		\vspace{0.2cm}
		\textbf{Density response:}
		After the quench into disorder, the random potential causes a spatially varying accumulation of phase ${\Delta\phi = Vt/\hbar}$, resulting in a velocity field according to ${\textbf{v}=\hbar / m \nabla \phi}$. We can only detect density variations once their size exceeds the resolution $\alpha=\SI{2.2}{\micro\meter}$ of our imaging system. Therefore, we are interested in the typical time $\trise$ after which the flow has traversed the distance $\alpha$. Thus, we estimate $\left<\left|\textbf{v}\right|\right>$ in order to be able to calculate ${\Delta s = 1/2 \left<\left|a\right|\right>t^2}$, where ${\left<\left|a\right|\right> = \mathrm{d}\!\left<\left|\textbf{v}\right|\right>\!/\mathrm{d}t = \left<\left|\nabla V\right|\right>\!/ m}$. Since the only relevant length and energy scale of the speckle in the imaging plane are given by $\avg{V}$ and the correlation length $\eta_{x,y}$, the magnitude of the mean speckle gradient must be proportional to $\avg{V}\!/\eta_{x,y}$. Indeed, a numerical simulation provides ${\left<(\nabla V)_x\right> = \left<(\nabla V)_y\right> = \avg{V}\!/ \eta_{x,y}}$, yielding ${\left<\left|\nabla V\right|\right> = \sqrt{2}\avg{V}\!/ \eta_{x,y}}$. This leads to the estimation ${\trise = \sqrt{2m\alpha/\!\left<\left|\nabla V\right|\right>} = \sqrt{\sqrt{2}m\alpha\eta_{x,y} /\!\avg{V}}}$.
			
		Once the speckle potential is rapidly extinguished, the density redistributes to adapt to the altered external potential. We assume that the typical speed of flow is given by $v$. We can only detect the redistribution as long as it occurs on a length scale larger than $\alpha$. This yields the estimation $\tdecay = \alpha / v$. Plugging in either the speed of sound $v_\mathrm{s}$, the average thermal velocity from the Maxwell-Boltzmann distribution ${\propto\sqrt{\boltzmann T / m}}$ (${T<\SI{100}{\nano\kelvin}}$), or the maximum velocity during a classical harmonic oscillation $R_x\omega_x$ in the dipole trapping potential yields values close to the observed times. $R_x$ denotes the Thomas-Fermi radius of the condensate along $x$. \linebreak
		
		\vspace{0.2cm}
		\textbf{Coherent hydrodynamic response:}
		Since we attribute the breakdown of hydrodynamics to the loss of phase coherence, it must be related to the spatially varying phase accumulation after the quench. The mean phase difference between two points in the BEC after time $t$ is $\left< \delta \phi \right> = \left<\Delta V\right>t/\hbar$, with the mean speckle potential difference ${\avg{\Delta V} = \avg{\left|V(\bv r) - V(\bv r')\right|}}$. From the numerical simulation we obtain ${\left<\Delta V\right>=\avg{V}}$, yielding $\tbreakdown=\hbar /\!\avg{V}$. In order to incorporate the differences in initial ($A_\mathrm{i}$) and final ($A_\mathrm{f}$) peak aspect ratio in $\tbreakdown$, we write $\tbreakdown=\hbar /\!\avg{V} \times \Delta A /A_\mathrm{i}$, where ${\Delta A=A_\mathrm{i} - A_\mathrm{f}}$.
		
		As the time scale of reoccurrence of hydrodynamics, we find $\treoccurrence = 2 R_y / v_\mathrm{s}=2\sqrt{2}/\omega_y$, where $R_y=\sqrt{2\mu/m}/\omega_y$ is the Thomas-Fermi radius along $y$. Similar as for $\tbreakdown$, we write $\treoccurrence = 2\sqrt{2}/\omega_y \times \left| \Delta A \right| /A_\mathrm{f}$.

	\subsection*{Calculation of the phase correlation length}
		We get the correlation length of the phase $\sigma_\phi$ by calculating the autocorrelation function
		\begin{align}
			\mathrm{AC}_\phi \left(\delta x, \delta y\right) &= \int f(x,y)f^*(x+\delta x,y+\delta y) \mathrm{d}x \mathrm{d}y
		\end{align}
		of the function ${f(x,y)=\sqrt{n(x,y)}\times\exp\left(\iu\phi(x,y,z=0)\right)}$, which is the product of the phase factor in the central plane of the cloud and the square root of the integrated column density, the latter of which accounts for the inhomogeneous density distribution. Both the phase factor and column density are obtained from the numerical simulation of the Gross-Pitaevskii equation. To reduce the computational effort, we limit the evaluation to the central $x$-$y$ plane at ${z=0}$. Hence, exploiting the Wiener–Khinchin theorem~\cite{wienerGeneralized1930}, the autocorrelation function is given by
		\begin{equation}
			\mathrm{AC}_\phi\left(\delta x, \delta y\right) = \mathcal{F}^{-1}\left(\mathrm{PSD}\right)\left(\delta x, \delta y\right),
			\label{eq:phase_autocorrelation_1}
		\end{equation}
		where
		\begin{equation}
			\mathrm{PSD}=\left|\mathcal{F}\left(\exp\left(\iu\phi(x,y,z=0)\right)\sqrt{n(x,y)}\right)(k_x,k_y)\right|^2
			\label{eq:phase_autocorrelation_2}
		\end{equation}
		is the power spectral density with the spatial frequencies $(k_x,k_y)$.
		We evaluate the Fourier transform $\mathcal{F}$ numerically using a fast Fourier transform algorithm. The phase-correlation length $\sigma_\phi$ in $x$- ($y$-) direction is defined as the distance along $\delta x$ ($\delta y$) across which $\mathrm{AC}_\phi$ drops to ${1-1/e\approx \SI{63}{\percent}}$ of its value at ${\delta x=\delta y=0}$.
	
	\subsection*{Supplemental information on the numerical simulations}
		\subsection{Spatial correlations of the simulated speckle potential}
\begin{figure}
				\includegraphics[width=1\columnwidth,clip]{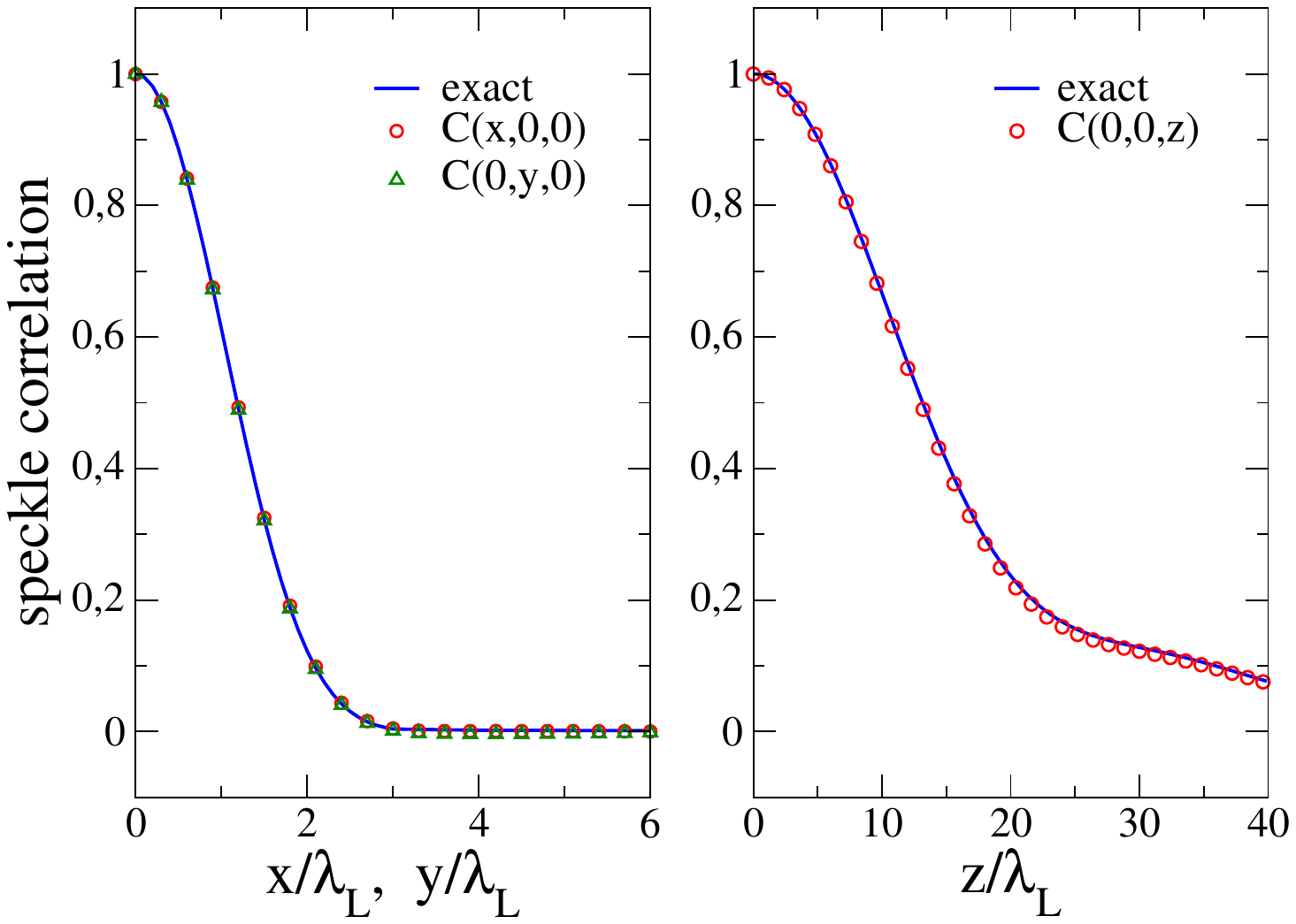}
				\caption{
					Spatial correlation function of the speckle potential along the $x$-, $y$-, and $z$-directions, orthogonal to the laser beam (left panel), and parallel to it (right panel). The distance is measured in units of the laser wavelength. The data symbols correspond to a numerical speckle generated on a cubic grid with sizes $N_x=N_y=N_z=1000$ and uniform spacing $\Delta x=\Delta y=\Delta z=0.3 \lambda_L$, while the solid lines refer to the correlations obtained from the direct integration of eq.~[7].}
				\label{fig:supp_figurespeckle}
			\end{figure}

			In Fig.~\ref{fig:supp_figurespeckle}, we display the spatial correlation function (see eq.~[6]) of the numerically generated disorder along the transverse (left panel) and longitudinal (right panel) directions. The plotted data correspond to a \textsl{single} disorder realization on a cubic grid.
			The results compare well with those obtained by a direct integration of eq.~[7], shown by the solid lines. 
			The correlation lengths obtained are the same as the ones for the experimentally employed disorder pattern.
	
	\subsubsection{Extracting the density response from numerical simulations}
		From the solution of the GP equation, we extract the 2D column density as $n_{2D}(\mathbf r_\perp,t)=\int n_{3D}(\mathbf r_\perp,z)\mathrm{d}z$. 
		Since $\sigma$ measures local density variations, it is much more sensitive to noise than other observables, like for instance the condensate widths.
		One source of noise is the finite spatial resolution of the imaging system, which tends to smear the microscopic density profile. 
		We account for this effect in a phenomenological way by convolving the numerical data with a 2D Gaussian function as
		\begin{equation}\nonumber
			n(\mathbf r_\perp,t)=\int n_{2D}(\mathbf r_\perp,t) \frac{1}{2\pi \alpha^2}  e^{-\frac{(x-x^\prime)^2+(y-y^\prime)^2}{2\alpha^2}} \mathrm{d}x^\prime \mathrm{d}y^\prime.
		\end{equation}
		For the Gaussian width we choose $\alpha=\SI{2.2}{\micro\meter}$.
		We then compute the time-resolved density response $\sigma$ and all the related observables (half-life period and steady-state value) from the convolved column density, following the same procedure used in the experiment. 
	
	\subsubsection{Influence of the grid size on the expansion dynamics}
\begin{figure}
			\includegraphics[width=1\columnwidth,clip]{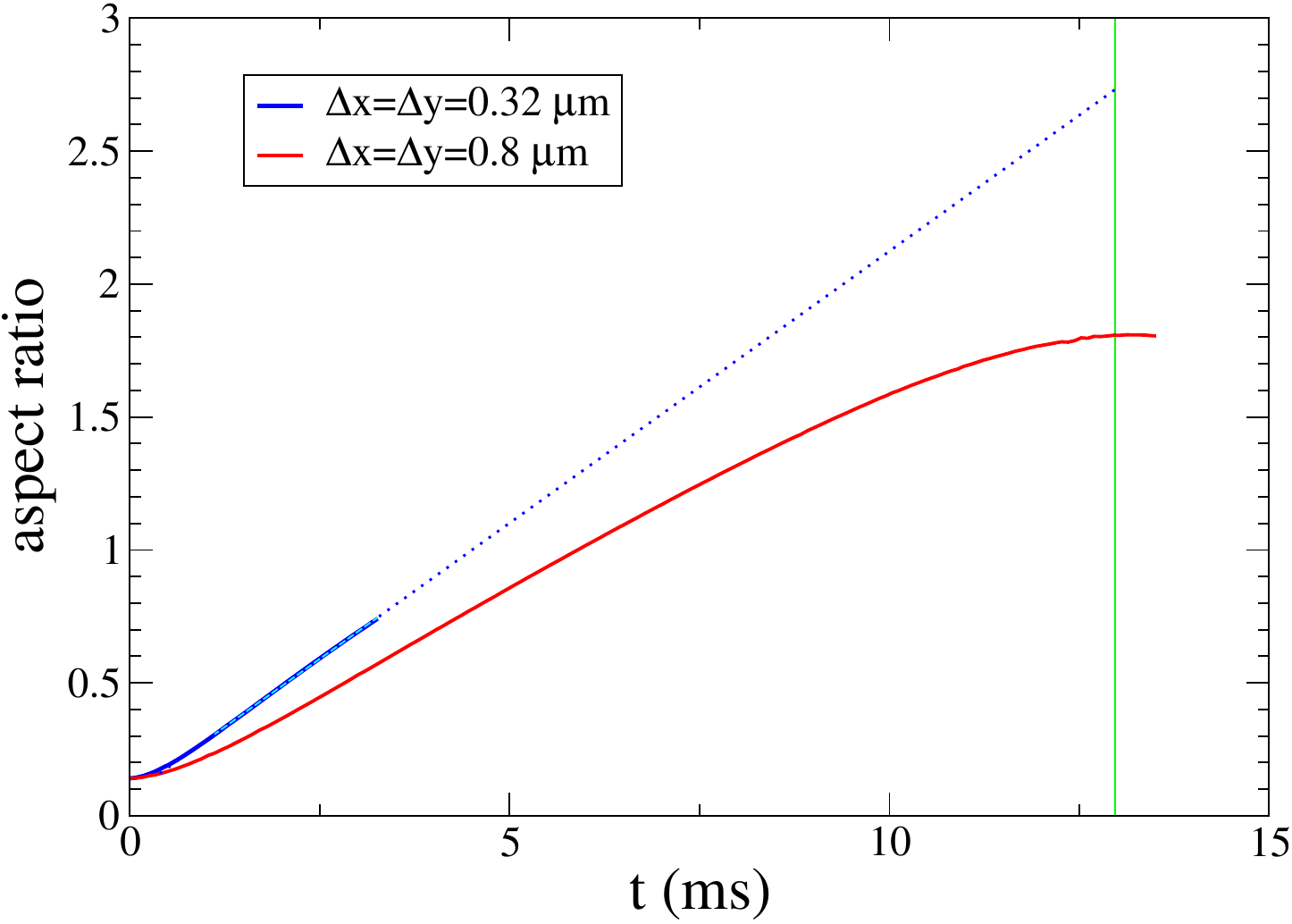}
			\caption{Aspect ratio as a function of time in the presence of the saddle point potential, calculated for $\langle V\rangle=\mu$ and large exposure time $\tau^\mathrm{on}=\SI{160.84}{\micro\second}$. The red solid curve is obtained by solving the GP equation with grid spacings $\Delta x=\Delta y=5\Delta z=\SI{0.8}{\micro\meter}$, yielding peak aspect ratio equal to $1.81$ for $t\simeq\SI{13}{\milli\second}$ (green vertical line). For large  $\tau^\mathrm{on}$  the cloud aspect ratio increases essentially linearly in time until the maximum is approached. The blue solid curve refers to the same calculation but with grid spacing $\Delta x=\Delta y=2\Delta z=\SI{0.32}{\micro\meter}$. Due to the much larger grid sizes, only a small part (approximately one fourth) of the simulation is numerically accessible. By linear extrapolation (blue dotted line), we obtain a peak aspect ratio in better agreement with the experimental data. 
			}
			\label{fig:supp_figureexpansion}
		\end{figure}

		Before the expansion, the trapped BEC is suddenly exposed to the speckle potential for a variable exposure time $\tau^\mathrm{on}$. We first solve the GP equation on a uniform grid to find how the condensate wavefunction is modified by the  disorder. We then use it as the initial state for the GP simulation of the subsequent hydrodynamic expansion. Since this kind of simulation is computationally daunting, especially in the presence of the remaining saddle-potential, we increase by a factor of five the grid spacings along the x and y directions. 
		This introduces an approximation in the numerics, in particular the peak aspect ratio for large $\tau^\mathrm{on}$ is underestimated, resulting in an apparent discrepancy with the experimental data. The effect is displayed in Fig.~\ref{fig:supp_figureexpansion}, where we show the calculated aspect ratio as a function of time for $\avg{V}=\mu$. The blue and red curves correspond to small and large in-plane grid spacings, respectively. Notice that the GP simulation for the fine mesh is limited to short times due to the large computational cost.
		We see from the coarse grid data that there is a wide interval of time where the aspect ratio grows approximately linearly. This suggests that a more quantitative agreement with the experimental data
		could be obtained by a linear extrapolation  from the numerical data obtained for small grid spacings, as shown in the figure by the blue dotted line.
	
	\subsection*{Additional data}
		\subsubsection*{Response to quenches into disorder for variable interaction strength.}
\begin{figure}[h!]
				\begin{center}
					\includegraphics[trim=5 18 25 32,clip]{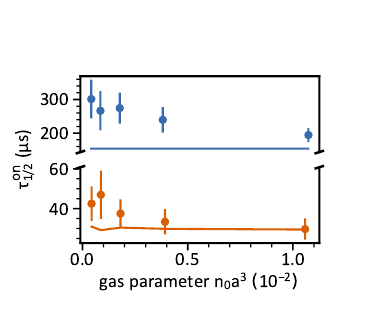}
					\caption{
						Response to quenches into disorder for variable interaction strength. Half-life periods for variable interaction strength and ${\avg{V}\!/\boltzmann=\SI{145}{\nano\kelvin}}$. The solid lines depict $\trise$ (blue) for the density response and $\tbreakdown$ (red) for the coherence.
					}
					\label{fig:supp_figure3}
				\end{center}
			\end{figure}

			With decreasing interaction strength but constant disorder strength ${\avg{V}/\boltzmann=\SI{145}{\nano\kelvin}}$, the response times of the density and coherence to quenches into disorder slightly increase (Fig.~\ref{fig:supp_figure3}).
	
		\subsubsection*{Losses.}
\begin{figure*}[h!]
				\includegraphics[trim=8 18 5 20,clip]{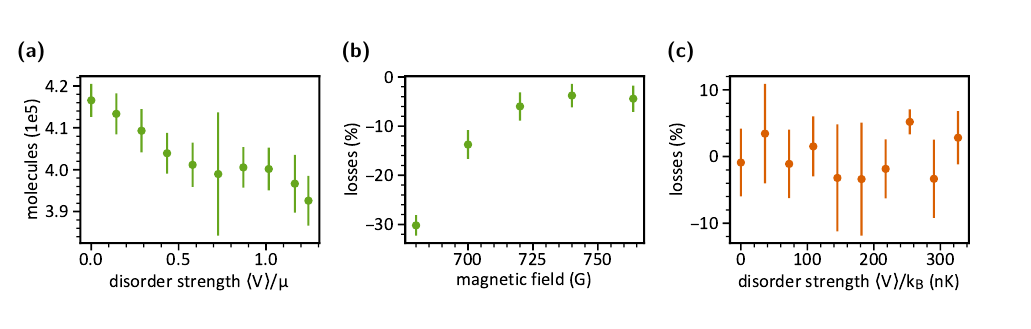}
				\caption{
					Losses.
					(a)~Molecule number of the measurement series probing the density response at \SI{763.6}{\gauss} for varying disorder strengths. Error bars are standard deviations of \num{5} repetitions with different disorder realizations.
					(b)~Relative losses for varying magnetic field and, thus, interaction strength at fixed disorder strength ${\avg{V}\!/\boltzmann=\SI{145}{\nano\kelvin}}$. Depicted is the ratio between molecule numbers after the quench out of and before the quench into disorder in the measurement series probing the density response.
					(c)~Relative losses during ${\tau_\mathrm{off}=\SI{150}{\milli\second}}$ after a quench out of disorder. Shown is the ratio between the molecule number after $\tau_\mathrm{off}$ and directly after the quench.
				}
				\label{fig:supp_figure2}
			\end{figure*}

			Introducing the speckle causes particle losses through several mechanisms. Since the mean disorder potential, chemical potential and optical trap depth are of similar magnitude, mere extrusion from the trap might occur. Besides, the presence of disorder can locally increase the density and thereby favor inelastic processes. Fig.~\ref{fig:supp_figure2}~(a) shows the molecule number of the measurement series probing the density response at \SI{763.6}{\gauss}, which decreases roughly linearly with disorder strength. For magnetic fields below \SI{720}{\gauss}, we observe increased losses (Fig.~\ref{fig:supp_figure2}~(b)) of up to \SI{30}{\percent}. This is, in part, caused by enhanced collisional relaxation of molecules into deeply bound states, an effect that leads to molecule losses and increases rapidly with decreasing scattering length~\cite{Petrov2004}. For quenches out of disorder, Fig.~\ref{fig:supp_figure2} shows the ratio between the molecule number after $\tau_\mathrm{off}$ and directly after the quench. The variations around zero and error bars reflect the typical particle number variation in our experiment.
		
	\bibliography{bibliography}

\end{document}